\documentclass[12pt]{article}

\usepackage{leftindex}
\usepackage{amsmath}
\usepackage{arxiv}
\usepackage{geometry}
\usepackage[utf8]{inputenc} % allow utf-8 input
\usepackage[T1]{fontenc}    % use 8-bit T1 fonts
\usepackage{hyperref}       % hyperlinks
\usepackage{url}            % simple URL typesetting
\usepackage{booktabs}       % professional-quality tables
\usepackage{amsfonts}       % blackboard math symbols
\usepackage{nicefrac}       % compact symbols for 1/2, etc.
\usepackage{microtype}      % microtypography
\usepackage{lipsum}		% Can be removed after putting your text content
\usepackage{graphicx}
\usepackage{natbib}
\usepackage{doi}
\usepackage{slashed}
\usepackage{cancel}
\usepackage{physics}
\usepackage{indentfirst}
\usepackage{multicol}
\usepackage{xcolor}
\numberwithin{equation}{section}
\usepackage{comment}
\usepackage{caption}
\newcommand{\tg}{\theta}
\newcommand{\tb}{\Bar{\theta}}

\title{Stellar Bounds on a Model with Photon-Photino Oscillation}

\author{ \href{https://orcid.org/0000-0003-3157-6005}{\includegraphics[scale=0.06]{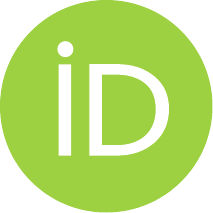}\hspace{1mm}Bernard T. de Menezes}%\thanks{Use footnote for providing further
        \\
	Brazilian Center for Research in Physics\\
	Rio de Janeiro, Brazil. \\
	\texttt{bernardmenezes123@gmail.com} \\
	\And
	\href{https://orcid.org/0000-0001-8310-518X}{\includegraphics[scale=0.06]{orcid.pdf}\hspace{1mm}José A. Helayël-Neto} \\
	Brazilian Center for Research in Physics\\
	Rio de Janeiro, Brazil. \\
	\texttt{josehelayel@gmail.com} \\
}

\renewcommand{\shorttitle}{}

\hypersetup{
pdftitle={A template for the arxiv style},
pdfsubject={q-bio.NC, q-bio.QM},
pdfauthor={David S.~Hippocampus, Elias D.~Striatum},
pdfkeywords={First keyword, Second keyword, More},
}

\begin{document}
\maketitle

\begin{abstract}
   In this paper, we pursue an investigation of the consequences of a mixing between supersymmetric partners - the photon and photino - analogous to the so-called Primakoff effect, but induced by a Lorentz-symmetry violating (LSV) fermionic-condensate background. In our framework, the LSV parameters are introduced as members of a non-dynamical superfield. As a consequence, we show that naturally there appears a mixing term between the gauge boson and the gaugino, which can be readily seen in the superspace/superfield approach. We inspect the kinetic photon-photino mixing matrix in the scenario of stellar physics which we apply our results to. Bounds on the strength of the fermionic LSV background are can be set by invoking the energy loss argument and the solar data we adopt.          
\end{abstract}

% keywords can be removed

\begin{multicols}{2}
\section{Introduction}

It is a common knowledge that Lorentz and CPT invariances are fundamental cornerstones of the Standard Model of Particle Physics (SM). However, at the quantum gravity scale (Planck scale), Lorentz invariance is expected to be violated\cite{KosteleckySpontaneousBreakLorentzString}, by virtue of the huge vacuum fluctuations, we refer to as spin foam. These fluctuations induce scenarios of physics beyond the Standard Model, as it is described, for example, by Superstring Theory \cite{KosteleckySpontaneousBreakLorentzString,KosteleckyGravitationalConstraints}. At this more fundamental stage, Supersymmetry (SUSY) could still be preserved; SUSY breaking is expected to be broken around the scale of grand-unification. In our endeavour, we assume that, at the scale where Lorentz symmetry violation takes place, SUSY is considered to be still a good symmetry. With this viewpoint, we approach LSV effects within a supersymmetric scenario  \cite{AlvaroSuSyMaxwellCScptViolation,Alvaro2004remarks}. 

All possible background tensor operators that could induce Lorentz- and CPT-violating effects are thoroughly contemplated in a broader framework, known as Standard Model Extension \cite{colladay1998lorentz}. Many formulations have been exhaustively discussed \cite{colladay1997cpt}, adopting a four-vector through a CPT-odd \cite{carroll1990limits} or a CPT-preserving \cite{belich2015aspects} term. Experimental tests may be carried out to impose constraints and limits on LSV effects induced in many sectors of the SM. The preservation of SUSY in the context of Lorentz violation was first proposed in \cite{bergerKostelecky2002supersymmetry}, with the LSV vector background coupled to the SUSY algebra. However, an alternative path \cite{AlvaroSuSyMaxwellCScptViolation} has been framed to preserve the SUSY algebra and introduce LSV via a superspace/superfield formalism. In this paper, we are aligned with the latter approach, and we recall, in Section (1), how to introduce LSV in N=1-SUSY superspace.

Our claim is that the LSV parameters in the supersymmetric scenario should be accommodated in a superfield which behaves as a background, that is, a non-dynamical entity. The bosonic LSV vector background gains an supersymmetric partner, i.e. an fermionic background that also breaks Lorentz symmetry. Recent works showed that this approach also induces the breaking of SUSY by giving the gauge-boson and the gaugino different masses in the spectrum \cite{AspectsofGaugeGauginoMixing}. The SUSY formalism allows for different fermionic background bilinear condensates that alters the gaugino propagator and also modifies the dispersion relations \cite{belich2015aspects}. Another striking feature is the appearance of a mixing term between the gauge-boson and the gaugino, induced by the fermionic background. This allows for a transition mechanism, analog to the so called Primakoff effect ,\cite{MixingofPhotonwithLowMassParticles,AxionLikeProduction}. 

Also, we would like to emphasize that our approach preserves gauge symmetry. Here, we aim at extending the Carroll-Field-Jackiew (CFJ) type term to include simple supersymmetry:
\vspace{.2cm}
\begin{equation}
\label{CFJterm}
    S_{CFJ} = \int d^4x \left[ \frac{1}{2}v_\mu A_\nu \Tilde{F}^{\mu\nu}\right],
\end{equation}
where $v^\mu$ is a LSV vector background and $A^\mu$ is the abelian gauge field.  In order to preserve gauge symmetry, it is imposed the following condition on the CFJ-background vector:
\begin{equation}
\label{CFJgaugeCondition}
    \partial_\mu v_\nu - \partial_\nu v_\mu =0.
\end{equation}
This can be seen by making a gauge transformation on the CFJ-term and, after an integration by parts, impose the condition \eqref{CFJgaugeCondition} to recover the gauge invariance. Our supersymmetric approach derives the relation \eqref{CFJgaugeCondition} as a by product of the bounds imposed on the LSV Superfield in order to obtain the CFJ-term \eqref{CFJterm}. The procedure is shown in more details in Section \ref{IntroducaoSusy}.

It is possible to show that, in the presence of a external magnetic field, a kinetic mixing between different bosonic fields is possible. This phenomena, i.e. the Primakoff effect, was studied in many different contexts, varying from the Axionic Electrodynamics \cite{AxionElectrodynamicsWilczek,AxionLikeProduction}, dark-photons \cite{MixingDarkphoton}, and gravitons \cite{MixingofPhotonwithLowMassParticles}. This feature is exploited in search for Axion like Particles (ALPs) in shining-through-walls experiments \cite{ShiningLightThroughWall}, for Solar ALPs in CAST experiment \cite{SolarAxions}, and for assignatures of dark matter ALPs in microwave cavity experiments, i.e. the ADMX experiment \cite{ADMXExperiment}. Nowadays, the Primakoff effect is regard, not only as mechanism to search for ALP's, but also as an access to the dark sector, i.e. a dark matter portal. 

In this paper, we show how the LSV fermionic background induces a supersymmetric version of the Primakoff Effect, but now mixing a fermion with a boson. Following \cite{AxionLikeProduction}, we contextualized the supersymmetric mixing on a stellar physics scenario, in particular using solar data. The production of photinos from a mixing process with photons can be considered a novel source of energy loss from a star, like the production of neutrinos from nuclear fusion at the star's core. This energy sink can be constrained by heliosismology and the neutrino's flux data, which imply a bound on the intensity of the fermionic LSV background \eqref{BoundFromEnergyLossArgument}. We discuss the details in Sections \ref{PhotonPhotinoRateProduction}, \ref{SunProfileAndPhotonsProduction} and \ref{PhotinoFluxAtEarth}.

We outline our paper as follows: in Section \ref{IntroducaoSusy}, we briefly review the superspace formalism and present the SUSY-LSV model we investigate. Next, in Section \ref{LinearizacaoEqsSUSYCFJ}, we linearize the field equations and show how we can compute the photon/photino probability transitions from the kinetic matrix that involves the photon and photino fields. To place the mixing in the context of stellar physics, in Section \ref{PhotonPhotinoRateProduction} we follow the path of \cite{AxionLikeProduction} and apply the density operator formalism to the mixed-field Hamiltonian shown in \eqref{MixingHamiltonian}. This allows us to compute the photon/photino production rate and the associated luminosity associated. This luminosity is calculated using solar models and data in Sections \ref{SunProfileAndPhotonsProduction} and \ref{PhotinoFluxAtEarth} . To impose bounds on the LSV fermionic background, we make use of the so-called energy loss argument \cite{StarsAsLaboratoryForFundamentalPhysics} in Section \ref{BoundFromEnergyLossArgument}. We cast our final comments and conclude the paper in Section \ref{Conclusions}.

\section{The Abelian Super-Gauge Model: the Warm-up}
\label{IntroducaoSusy}

In this Section, we shall briefly review the $N=1$-SUSY Abelian model  \cite{bilal2001introduction}. We start off with the chiral/anti-chiral multiplets $(0,1/2)\oplus (-1/2,0)$, defined by the conditions:
\begin{equation}
\label{ChiralityCondition}
    \Bar{\mathcal{D}}_{\Dot{\alpha}}\phi = \mathcal{D}_\alpha \Bar{\phi}=0,
\end{equation}
with the SUSY covariant derivatives defined by:
\begin{equation}
\label{SUSYConvariantDerivartive}
    \{\mathcal{D}_\alpha, \bar{\mathcal{D}}_{\dot{\beta}}\} = -2i \sigma^\mu_{\alpha\dot{\beta}}\partial_\mu.
\end{equation}

As a second step, we present the vector multiplet $(1,1/2)\oplus (-1,-1/2)$ in the so called Wess-Zumino supergauge:
\begin{equation}
\label{VectorFieldWZgauge}
    V_{\textrm{WZ}}=\theta \sigma^\mu \bar{\theta} A_\mu (x)+ i \theta^2 \bar{\theta}\bar{\lambda}(x) +i\bar{\theta}^2\theta \lambda (x) + \theta^2 \bar{\theta}^2 D(x).
\end{equation}
Its gauge transformation reads as given below:
\begin{equation}
\label{SuperGaugeTransformations}
    V'=V + \phi +\phi^\dagger,
\end{equation}
which can be shown to yield the usual gauge transformation of the vector potential:
\begin{equation}
    A_\mu'=A_\mu + \partial_\mu\alpha(x).
\end{equation}
In order to build up the Abelian superfield kinetic term, we act with the SUSY covariant derivatives on the vector multiplet \eqref{VectorFieldWZgauge} as:
\begin{equation}
\label{SuperAbelianFieldStrenght}
    W_\alpha=\frac{g}{2}\bar{\mathcal{D}}^2\mathcal{D}_\alpha V_{WZ}, \ \ \ \  \bar{W}_{\dot{\alpha}}=\frac{g}{2}\mathcal{D}^2\bar{\mathcal{D}}_{\dot{\alpha}} V_{WZ}.
\end{equation}
Using the super-chirality condition \eqref{ChiralityCondition}, we can shown that \eqref{SuperAbelianFieldStrenght} is invariant under the abelian super-gauge transformations \eqref{SuperGaugeTransformations}. With all set, we construct the following SUSY-Maxwell action:
\begin{equation}
\label{SUSYMaxwell}
    \begin{split}
    \mathcal{S}^{\textrm{SUSY}}_{\textrm{Maxwell}}&=\frac{-1}{4g^2}\int d^4x\left(\int d^2 \theta \ W^\alpha W_\alpha+\int d^2\Bar{\theta} \ \Bar{W}_{\dot{\alpha}}\Bar{W}^{\dot{\alpha}}\right)\\ 
    &=\int d^4x \left(-\frac{i}{2}\Lambda \slashed{\partial}\bar{\Lambda} + \frac{1}{2}D^2 -\frac{1}{4}F^{\mu\nu}F_{\mu\nu}\right),
    \end{split}
\end{equation}
with the spinors in the 4-component notation,
\begin{equation}
    \Lambda=\begin{pmatrix}
        \lambda_\alpha \\
        \Bar{\lambda}^{\dot{\alpha}}
    \end{pmatrix},
\end{equation}
and the abelian field strength as
\begin{equation}
    F^{\mu\nu}=\partial^\mu A^\nu - \partial^\nu A^\mu.
\end{equation}
Here, we emphasize that D is just an auxiliary field which we eliminate with the help of the Euler-Lagrange equations of the theory.

Now, we propose the supersymmetric version of CFJ-term as
\begin{equation}
\label{SuperfieldCFJAction}
    \mathcal{S}_{\textrm{CFJ}}^{\textrm{SUSY}}= \int d^4xd^4\theta\{W^{\alpha} (D_\alpha V)S+ c.c.\},
\end{equation}
where c.c. denotes complex conjugation. We have introduced the breaking of Lorentz symmetry in the chiral superfield $S(x,\theta)$:
\begin{equation}
    \begin{split}
    S(x,\theta) &= s(x)+i\tg \sigma^\mu \tb\partial_\mu s(x)-\frac{1}{4}\tg^2 \tb^2\partial^2 s(x) +\sqrt{2}\tg \psi(x)\\
    &+\frac{i}{\sqrt{2}}\tb^2 \tb \Bar{\sigma}^\mu\partial_\mu \psi (x) + \tg^2f(x),  
    \end{split}
\end{equation}
with $\bar{\mathcal{D}}_{\Dot{\alpha}}S(x,\theta)=0$. This superfield is neutral under the gauge group and its canonical mass dimension is 0. The application of the SUSY covariant derivative in the vector multiplet is given by
\begin{equation}
    \begin{split}
    \mathcal{D}_\alpha V_{\textrm{WZ}} &= (\sigma^\mu \tb)_\alpha A_\mu +2\theta_\alpha \bar{\theta}\bar{\lambda} +\bar{\theta}^2\lambda_\alpha + 2\theta_\alpha \tb^2 D  \\
    & + \tb^2 (\sigma^{\mu\nu}\tg)_\alpha F_{\mu\nu}  
     - \frac{i}{2} \tg^2\tb^2 (\slashed{\partial} \bar{\lambda})_\alpha,
     \end{split}
\end{equation}
and $W_\alpha$ as in \eqref{SuperAbelianFieldStrenght}. Thus, projecting out in field components the action \eqref{SuperfieldCFJAction}, we find:
\begin{equation}
    \begin{split}
             &\mathcal{S}_{\textrm{CFJ}}^{\textrm{SUSY}} =\int d^4x \bigg\{   +4\Re{s}i\Bar{\lambda} \Bar{\sigma}^\mu \partial_\mu \lambda 
            +8\Re{s}(D)^2\\ 
            & -\Re{s}(F_{\mu\nu})^2-2\epsilon^{\mu\nu\lambda\kappa}(\partial_\mu\Im(s)) A_\nu \partial_\lambda A_\kappa \\
            &-\sqrt{2}(\lambda \sigma^{\mu\nu}\psi-\bar{\lambda} \bar{\sigma}^{\mu\nu}\bar{\psi})F_{\mu\nu} 
            -2\sqrt{2}(\lambda\psi+\bar{\lambda}\bar{\psi})D \\
            &+ (\lambda)^2f+(\bar{\lambda})^2 f^\dagger\bigg\}.
        \end{split}
\end{equation}
To obtain the CFJ-term, we impose under $S(x,\theta)$ the following boundaries conditions:
\begin{equation}
\label{SUSYConditions}
    \begin{split}
    &\Re{s}=\frac{s+s^\dagger}{2}=0 \rightarrow s=-s^\dagger \ \ \textrm{and} \\ 
    &\ \ \Im{s}=\varphi = \frac{v_\mu x^\mu}{4},
    \end{split}
\end{equation}
such that,
\begin{equation}
    \partial_\mu \Im{s}=\partial_\mu \varphi=\frac{v_\mu}{4}.
\end{equation}
Notice that SUSY already provides the 4-curl condition that preserves the gauge symmetry of the CFJ term:
\begin{equation}
    v_\mu \propto \partial_\mu \varphi \ \ \rightarrow \ \ \partial_\mu v_\nu - \partial_\nu v_\mu =0.
\end{equation}
Substituting the SUSY conditions \eqref{SUSYConditions} and adding the action \eqref{SUSYMaxwell}, we finally find the supersymmetric version of the CFJ-Electrodynamics in 4-components notation:
\begin{equation}
\label{SUSYCFJmodel}
        \begin{split}
        &S^{\textrm{CFJ-ED}}_{\textrm{SUSY}}=\int d^4x \bigg\{\underbrace{-\frac{(F_{\mu\nu})^2}{4}}_{\textrm{Gauge term}}-\underbrace{\frac{1}{2}\varepsilon^{\mu\nu\lambda\kappa}v_\mu A_\nu\partial_\lambda A_\kappa}_{\textrm{CFJ term}}\\
        &-\frac{i}{2}\bar{\Lambda}\slashed{\partial}\Lambda -\frac{R_\mu}{4}\bar{\Lambda}\gamma^\mu \gamma_5 \Lambda 
         +\sqrt{2}(\bar{\Lambda}\Sigma^{\mu\nu}\gamma_5\Psi)F_{\mu\nu} \\
         &+ M_1\bar{\Lambda}\Lambda -iM_2\bar{\Lambda}\gamma_5\Lambda \bigg\},
    \end{split}
\end{equation}
where we defined the quantities:
\begin{equation}
    \begin{split}
    & R_\mu=v_\mu+\bar{\Psi}\gamma_\mu \gamma_5 \Psi, \ \  M_1= \Re{f} + \frac{\bar{\Psi}\Psi}{4}, \\
    & M_2=\Im{f}+\frac{i}{4}\bar{\Psi}\gamma_5 \Psi.
    \end{split}
\end{equation}
Now, we proceed to derive a linearized version of the SUSY-CFJ model \eqref{SUSYCFJmodel} and the kinetic transition matrix.

\section{Linearization of the SUSY-CFJ Model in Presence of the Plasma Medium}
\label{LinearizacaoEqsSUSYCFJ}
In the past section, we have shown how a supersymmetric version of the CFJ-Electrodynamics generates a mixing term of the type $-\Bar{\Psi}\Sigma^{\mu\nu}\gamma_5\Lambda F_{\mu\nu}$. Now, we wish to derive the kinetic mixing matrix of the model and demonstrate the probabilities transitions associated with oscillating states. For simplicity, we consider here $M_1=M_2=\Bar{\Psi}\gamma^\mu \gamma_5 \Psi=0$. In this context, the Lagrangian for the SUSY-CFJ Abelian model, in the presence of a plasma source $k^2=\omega^2 - |\Vec{k}|^2 - \omega_p^2$ \ , in units $\varepsilon_0=\mu_0=1$, has the form:

\begin{equation}
\label{LagrangianAbelianCFJ}
    \begin{split}
        \mathcal{L} &=-\frac{F_{\mu\nu}^2}{4}-\frac{1}{2}\varepsilon^{\mu\nu\alpha\beta}v_\mu A_\nu\partial_\alpha A_\beta  -\frac{i}{2}\bar{\Lambda}\slashed{\partial}\Lambda \\
        &-\frac{1}{4}\bar{\Lambda}\slashed{v} \gamma_5 \Lambda  
        +2\sqrt{2}(\bar{\Lambda}\Sigma^{\mu\nu}\gamma_5\Psi)\partial_\mu A_\nu. 
    \end{split}
\end{equation}
Here, we use the $\gamma$-matrices in Weyl's representation:

\begin{equation}
    \begin{split}
    &\gamma^\mu = \left(\begin{array}{cc}
        0 & \sigma^\mu \\
        \bar{\sigma}^\mu & 0
    \end{array}\right),  \ \ \ \gamma^5=\begin{pmatrix}
        -\textbf{1}_2 & 0 \\
        0 & \textbf{1}_2
    \end{pmatrix}, \\
    &\ \ \ C=\begin{pmatrix}
       -\varepsilon & 0 \\
        0 & \varepsilon
    \end{pmatrix}, \ \ \ \begin{cases}
        \sigma^\mu =& (\textbf{1}_2, \sigma^i), \\
        \bar{\sigma}^\mu =& (\textbf{1}_2, -\sigma^i),
    \end{cases}
    \end{split}
\end{equation}
with $\sigma^i$ the Pauli's matrices and $\varepsilon=i\sigma^2$. In this representation, the Majorana spinors are written as
\begin{equation}
    \Lambda = C\Bar{\Lambda}^t= \begin{pmatrix}
        \lambda_\alpha \\
        \Bar{\lambda}^{\Dot{\alpha}}
    \end{pmatrix}, \ \ \ \Bar{\Lambda} = \begin{pmatrix}
         \lambda^\alpha , \ \ \Bar{\lambda}_{\Dot{\alpha}}
    \end{pmatrix}.
\end{equation}
The Lorentz generators with the chirality matrix $\gamma_5$ are given by
\begin{equation}
    \Sigma^{\mu\nu}\gamma_5 =i\begin{pmatrix}
        -(\sigma^{\mu\mu})_\beta^{\ \ \alpha} & 0\\
        0 & (\Bar{\sigma}^{\mu\mu})^{\Dot{\beta}}_{\ \ \Dot{\alpha}}
    \end{pmatrix}.
\end{equation}
Since we are working with Majorana spinors, is more easy to work in the 2 components notation. Assuming the Lorenz gauge ($\partial_\mu A^\mu=0$), in 4-momentum space, the Lagrangian \eqref{LagrangianAbelianCFJ} can be cast as
\begin{equation}
\label{4MomentumLAgrangian}
    \begin{split}
        \mathcal{L} &= -\frac{k^2}{2}A^\mu A_\mu +\frac{i}{2}\varepsilon^{\mu\nu\alpha\beta}v_\mu A_\nu k_\alpha A_\beta - \Bar{\lambda} \Bar{\sigma}^\mu k_\mu \lambda \\
        & +\frac{1}{2}\Bar{\lambda}\Bar{\sigma}^\mu v_\mu \lambda + 2\sqrt{2}(\lambda \sigma^{\mu\nu}\psi - \Bar{\lambda}\Bar{\sigma}^{\mu\nu}\Bar{\psi})k_\mu A_\nu. 
    \end{split}
\end{equation}

In order to show the kinetic mixing, we must evaluate the Lagrangian \eqref{4MomentumLAgrangian} in a linearized form, i.e. $\omega^2 - k_z^2=(\omega+k_z)(\omega - k_z)\approx 2\omega (\omega-k_z)$. For the bosonic sector, we consider the photon in the polarization basis $A_\pm = (A^x \pm i A^y)/\sqrt{2}$, the Coulomb gauge $\Vec{k}\cdot \Vec{A}=0$, and the bosonic background as $v^\mu=(0,0,0,v_z)$. In the fermionic sector, we consider the photino normalized as $\lambda = \sqrt{\omega}\lambda(k)$. 
%Also, since $\Bar{\Psi}\Psi=\Bar{\Psi}\gamma_5\Psi=0$, we have that the fermionic background components are equal, i.e. $\psi_1=\psi_2$. 
Taking into account all this context, the Lagrangian can be cast in the form:

\begin{equation}
\label{KineticMixingMatrix}
    \begin{split}
        \mathcal{L} &= \omega^2 A_+ A_- -\omega^2 |\lambda_1|^2 -\omega \lambda_2^* (\omega - k_z -\Delta_R^-)\lambda_2 \\
        &+\omega \begin{pmatrix}
        A_+ & \lambda_1^*  
        \end{pmatrix}\underbrace{\begin{pmatrix}
        -k_z-\Delta_v & \chi \\
       \chi^* & -k_z-\Delta_R^+ 
        \end{pmatrix}}_{=K_c}\begin{pmatrix}
        A_- \\
        \lambda_1
    \end{pmatrix},
    \end{split}
\end{equation}
where we defined the parameters: 
\begin{equation}
    \begin{split}
        &\chi=4\sqrt{\omega}\psi_1, \\
        &\Delta_v=\frac{\omega_p^2}{2\omega}-v_z/2, \\
        &\Delta_R^\pm= \mp \frac{v_z}{2}, \\
        & \omega_p^2= \frac{n_e e^2}{m_e},
    \end{split}
\end{equation}
and the last parameter being the so called plasma frenquency of the medium.

We can see from the kinetic matrix $K_c$ that the fields cannot be simultaneously eigenstates of 3-momentum and energy. Here, we choose the latter case but it is possible to derive a similar result assuming the first case. Varying the action with respect to the fields, we find the field equations, in matrix form:
\begin{equation}
    \begin{split}
        &\begin{pmatrix}
            \omega- k_z - \Delta_v & \chi & 0\\
               \chi^* &-\omega-k_z-\Delta_R^+ & 0 \\
                0 & 0 &-\omega +k_z-\Delta_R^- 
                \end{pmatrix} \\
                & \times \begin{pmatrix}
                A_- \\
                 \lambda_1\\
                \lambda_2
                 \end{pmatrix}=0_{3\times 1}.
    \end{split}
\end{equation}

Now, we proceed to derive the transition probabilities for the  photon/photino system through linearized propagators in configuration space. We can obtain the propagators in 3-momentum space by inverting the kinetic matrix $K_c$:
\begin{equation}
    K_c^{-1}= 
   \frac{1}{D}\begin{pmatrix}
        -k_z - \Delta_R & -\chi \\
        -\chi^* & -k_z -\Delta_\nu -\frac{2|\chi|^2}{(k_z+\Delta_R)}
    \end{pmatrix}
\end{equation}
where we defined the quantity $D = k_z^2+k_z(\Delta_R^++\Delta_v)+\Delta_v\Delta_R^+ + |\chi|^2$. The elements of $K_c^{-1}$ are the correlation functions in 3-momentum space, i.e. $\bra{\phi_b}U(k_z)\ket{\phi_a}$, with $\phi_a=A, \lambda$. 

The off-diagonal elements in configuration space gives us the transition amplitudes. The photino/photon transition amplitude is given by:
\begin{equation}
\label{CorrelationPhotinoPhoton1}
    \begin{split}
    &G_{\lambda \rightarrow A}(\omega,z) = \bra{A(0)}\ket{\lambda(z)} \\
    &=-\chi \int \frac{dk_z}{2\pi}\frac{e^{ik_z z}}{(k_z- \alpha)(k_z-\beta)} \\ 
    &= -\frac{i\chi}{2\Delta} \int\frac{dk_z}{2\pi}e^{ik_z z}\left[\frac{1}{k_z -\alpha}-\frac{1}{k_z -\beta}\right] \\
        &= -\frac{i\chi}{2\Delta}\left[e^{-iz\left(\Delta_R + \frac{|\chi|^2}{\Delta_\nu - \Delta_R}\right)}-e^{-iz\left(\Delta_\nu - \frac{|\chi|^2}{\Delta_\nu - \Delta_R}\right)}\right],
    \end{split}
\end{equation}
with $\Delta= (\Delta_\nu - \Delta_R)/2 - |\chi|^2/(\Delta_\nu - \Delta_R)$ and the denominator's roots are given by $\alpha=-\Delta_R - |\chi|^2/(\Delta_v - \Delta_R)$ and $\beta= -\Delta_v+|\chi|^2/(\Delta_v - \Delta_R)$.  We can simplify some expressions through Taylor expanding the Grassimanian functions:
\begin{equation}
    \begin{split}
        -\frac{i\chi}{2\Delta} &= -\frac{i\chi}{(\Delta_v -\Delta_R)}\left(1-\frac{2|\chi|^2}{(\Delta_v-\Delta_R)^2}\right)^{-1} \\
        &=-\frac{i\chi}{(\Delta_v -\Delta_R)}\left(1+\frac{2|\chi|^2}{(\Delta_v-\Delta_R)^2}\right)\\
        &= \frac{-i\chi}{(\Delta_v-\Delta_R)}.
    \end{split}
\end{equation}
We can also simplify the Grassimanian contribution to the complex exponential as:
\begin{equation}
\label{GrassimanianSimplifications}
    \begin{split}
        \frac{-i\chi}{(\Delta_v-\Delta_R)}e^{\pm\frac{iz|\chi|^2}{\Delta_\nu - \Delta_R}} &= \frac{-i\chi}{(\Delta_v-\Delta_R)} \left(1\pm\frac{iz|\chi|^2}{\Delta_\nu - \Delta_R}\right)\\ 
        &= \frac{-i\chi}{(\Delta_v-\Delta_R)}.
    \end{split}
\end{equation}
Therefore the correlation function assumes the simplified form:
\begin{equation}
\label{CorrelationFunction1}
    G_{A\rightarrow \lambda}(\omega,z) = \frac{i\chi}{\Delta_v - \Delta_R}(e^{-iz\Delta_v}-e^{-iz\Delta_R}).
\end{equation}

The squared modulus gives us the probability amplitude:
\begin{equation}
    \boxed{P_{A\rightarrow \lambda}(\omega,z) = \frac{|\chi|^2}{\Delta_{osc}^2}\sin^2{(\Delta_{osc}z)}}
\end{equation}
where $\Delta_{osc}=(\Delta_v - \Delta_R)/2$.  As can be seen, the oscillation length $l_{osc}=2\pi/\Delta_{osc}$ (which is a measure of the scale where the oscillation can be observed) does not depend on the fermionic background intensity.

\section{Photon/Photino Production Rate}
\label{PhotonPhotinoRateProduction}

Assuming energy eigenstates, we can obtain the "Schrödinger like" oscillation equation in 3-momentum space as:
\begin{equation}
\label{MixingHamiltonian}
    i\partial_z \begin{pmatrix}
        A \\
        \lambda
    \end{pmatrix} = \underbrace{\begin{pmatrix}
        \Delta_v & -\chi \\
        -\chi^* & \Delta_R
    \end{pmatrix}}_{=H_I}\begin{pmatrix}
        A \\
        \lambda
    \end{pmatrix}
\end{equation}
In order to calculate the production rate of photons/photinos due to particle oscillations, we must consider a density matrix formalism.  We start by writing the interaction matrix $H_I$ as
\begin{equation}
    H_I=\begin{pmatrix}
        \Delta_v & -\chi \\
        -\chi^* & \Delta_R
    \end{pmatrix}= \frac{\Delta_v + \Delta_R}{2} \textbf{1}_2 + \begin{pmatrix}
        +\Delta \omega & -\chi \\
        -\chi^* & -\Delta \omega
    \end{pmatrix},
\end{equation}
with $\Delta \omega=(\Delta_v-\Delta_R)/2$. We consider an ensemble of photinos/photons that interacts, throght the presence of the fermionic background, with the stellar plasma without scattering of 3-momentum. The density matrix in thermal equilibrium is 
\begin{equation}
    \rho_T = \begin{pmatrix}
        f_T & 0 \\
        0 & 0
    \end{pmatrix},
\end{equation}
with the thermal equilibrium distribution of the photon/photinos given by $f_T=(e^{\omega/T}-\eta)^{-1}$., with $\eta=1$ for photons (bosons) and $\eta=-1$ for photinos (fermions). Considering that the stellar plasma background can interact with the photon/photino field and emit or absorb quanta of energy, we have the time evolution equation:
\begin{equation}
    \dot{\rho}= -[\Omega, \rho] + \frac{1}{2}\{G_{prod},1+\eta \rho\}-\frac{1}{2}\{G_{abs},\rho\},
\end{equation}
where we have defined the matrices:
\begin{equation}
    G_{\textrm{prod/abs}}=\begin{pmatrix}
        \Gamma_{\textrm{prod/abs}} & 0 \\
        0 & 0 
    \end{pmatrix}.
\end{equation}
Now, we consider a small deviation from thermal equilibrium by a perturbation:
\begin{equation}
    \rho=\rho_T + \delta \rho =\begin{pmatrix}
        f_T & 0 \\
        0 & 0 
    \end{pmatrix} + \begin{pmatrix}
        n_a & g \\
        g^* & n_e
    \end{pmatrix}.
\end{equation}
The time evolution equations become:
\begin{equation}
    \dot{\rho} = -i[H_I,\rho] -\frac{1}{2}\{G, \delta \rho\},
\end{equation}
with $G=\textrm{diag}(\Gamma,0)$ and $\Gamma=(1-\eta e^{- \omega/T})\Gamma_{abs}$. We find the following equations:

\begin{equation}
    \begin{split}
         &\dot{n_a} = -2\Im(\chi g^*) - \Gamma n_a, \\
         & \dot{n_e} = 2\Im(g^*\chi), \\
          & \dot{g} = -g(2i\Delta \omega + \Gamma/2)-i\chi(f_T + n_a - n_e). \\
    \end{split}
\end{equation}
Assuming that the excitation's do not deviate too much from the thermal equilibrium, i.e, $n_a,n_e << f_T$, then
\begin{equation}
    \dot{g}= -g(2i\Delta \omega + \Gamma/2)-i\chi f_T.
\end{equation}
Considering the initial condition $g(0)=0$, we find the solution for the decoherence:
\begin{equation}
    g(t)= -\frac{1-e^{-i(2\Delta \omega+\Gamma/2)t}}{2\Delta \omega-i\Gamma/2}\chi f_T.
\end{equation}
After an initial transient, it approximates to the steady-state solution as
\begin{equation}
    g(\infty)= -\frac{2\Delta \omega+i\Gamma/2}{(2\Delta \omega)^2 +\Gamma^2/4}\chi f_T.
\end{equation}
Inserting this solution into the equations for the evolution of $n_e$, we finally find the production rate of photons/photinos, with frequency $\omega$, as
\begin{equation}
    \label{ProductionRate}
    \boxed{\Gamma^{\textrm{prod}}_\eta \equiv\dot{n_e} = \frac{\Gamma |\chi|^2}{(\Delta_v-\Delta_R^+)^2+\Gamma^2/4}\frac{1}{e^{\omega/T}-\eta}}
\end{equation}

\section{The Sun's Profile and the Photon Production}
\label{SunProfileAndPhotonsProduction}

Now, we consider the situation where we have a production of photinos at the sun's core. We assume the photinos as sterile particles with no interaction that escapes from the solar domain, and therefore constitute a source for the energy loss of the star. Using the available data from the Sun \cite{bahcall2004we,asplund2009chemical,krief2016solar}, we wish to impose bounds on the fermionic background intensity.

The sun physics is model by a $n=3$ polytropic model \cite{kippenhahn1990stellar}. Solving the structure equations with this model gives us the following profile for the Sun:
\begin{equation}
    \begin{split}
        \rho(r) &= \rho_0 \  \theta^3(r/a), \\
        T(r) &= T_0 \ \theta(r/a).
    \end{split}
\end{equation}
with the central quantities given by,
\begin{equation}
    \begin{split}
        \rho_0  &= 76.35 \ g/cm^3; \\
        T_0 &= 1632.49 \ eV.
    \end{split}
\end{equation}

\begin{center}
    \includegraphics[scale=0.5]{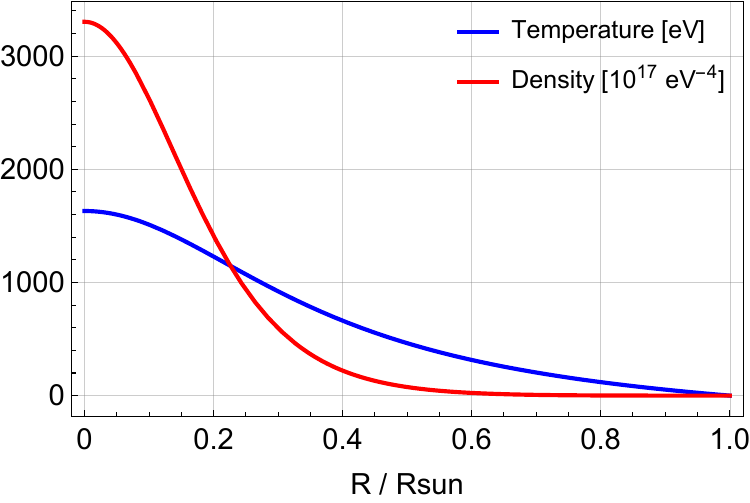}
    \captionof{figure}{Temperature (blue) and density (red) profiles.}
    \label{fig:fig1}
\end{center}
   
The function $\theta(r/a)$ is a numerical solution of the so-called Lane-Emden $n=3$ equation. This is derived from the structure equations proposing the polytropic ansatz where the pressure has the form $P(r) \propto \rho(r)^{1+1/n}$. We consider the sun's radius $R_\odot=6.96 \times 10^8 \ m$ and mass $M_\odot=1.99 \times 10^{30} \ kg$ as input data for the model and we defined $a=b/R_\odot$, such that $b=6.897$ and $\theta(b)=0$. For the sun chemical composition, we considered the abundances in \cite{asplund2009chemical}. The radial dependency of the density is also transmitted to the electron's number density, i.e. $n_e=n_e(r)$, which also implies a radial dependency on the plasma frequency $\omega_p^2=\omega^2_p(r)$.

From the thermal equilibrium condition of the star, we can see that the production and absorption rates are correlated. For the typical temperatures of the sun, the main sources of opacity (photon's absorption processes) are inverse bremsstrahlung at low and mid energies and Thomson's scattering at high energies \cite{MixingDarkphoton}:  

\begin{equation}
\label{EmissionRatePhotons}
    \begin{split}
        \Gamma_{abs}(r,\omega) =& \frac{64 \pi^3 \alpha^3}{3\omega^3 m_e^{3/2}\sqrt{2\pi T(r)}}\frac{\rho^2(r)}{\mu_e m_h^2} \sum_j Z_j^2 \frac{X_j}{A_j}\\
        & \times F\left(\frac{\omega}{T(r)}\right) 
        + \frac{8\pi\alpha^2}{3m_e^2}\frac{\rho(r)}{\mu_em_h},
    \end{split}
\end{equation}
where $m_h$ is an atomic mass unit, $\mu_e$ is the electron's mean weight and $X_j$ is the abundance of the j-th chemical species with proton number $Z_j$. For a pedagogical review of the derivation of the physical processes that influence sun's opacity, we recommend the reader \cite{IckoIbenStarsEvolution}. The function
\begin{equation}
    F(x) = K_0(x/2) \sinh{(x/2)},
\end{equation}
where $K_0$ is a Bessel function of the second kind, is an analytical expression obtained in the regime where screening effects of the ion's nucleus are neglected. Numerical estimations show that, in the energy regime of interest, the presence of screening reduces the emission by a few percentage and it is also compensated by other effects. Therefore, they are discarded for the purposes of this work.

At last, we would like to mention that, in expression \eqref{EmissionRatePhotons}, many other processes that contribute to the sun's opacity have been left aside, such as bound-free and bound-bound transitions, Pauli blocking, Sommerfeld enhancement and Coulomb scattering. They contribute much less to the photon's production rate \eqref{EmissionRatePhotons} in the Sun, so they can be regarded as precision refinements to our calculations. 

\section{Flux of Photinos on Earth}
\label{PhotinoFluxAtEarth}

Now, with the phenomenological scenario into consideration, we can compute the luminosity produced by the mixing process \eqref{ProductionRate}. The flux of photinos on Earth is given by
\vspace{.3cm}
\begin{equation}
    \dv{N_e}{t} = \frac{g=2}{4\pi D^2_{\odot}}\int d^3r \int \frac{d^3k}{(2\pi)^3} \Gamma^{\textrm{prod}}_{+1},
\end{equation}
where $D_\odot$ is the Sun-Earth distance and $g=2$ is the two polarization states of the photino. Assuming relativistic states $\omega \approx |\va{k}|$, we can insert a delta function in the photino 3-momentum integral $\delta(|\va{k}|- \omega)$. Therefore, the photino differential spectrum measured on Earth is:
\vspace{.3cm}
\begin{equation}
\label{DifferentialSpectrum}
    \begin{split}
        &\dv{\Phi_e}{\omega} =\frac{(2)(4\pi)^2\omega^2}{4\pi D^2_{\odot} (2\pi)^3} \int^{R_s}_0 r^2  \Gamma^{\textrm{prod}}_{+1} dr \\
        &= \frac{16}{(\pi^2 D^2_{\odot}) }  \int^{R_s}_0 r^2 \frac{\Gamma(r,\omega) |\psi|^2}{\left(\frac{\omega_p^2(r)}{2\omega}\right)^2+\Gamma(r,\omega)^2/4}\frac{\omega^3}{(e^{\omega/T(r)}-1)}.
    \end{split}
\end{equation}
Integrating \eqref{DifferentialSpectrum} in the radial variable and in the energies allows us to obtain the total luminosity of photinos produced at the sun:
\begin{equation}
    L_{\Lambda} = 4.73 \times 10^{30} |\psi|^2 L_{\odot}.
\end{equation}

\section{Bounds from the Energy Loss Argument}
\label{BoundFromEnergyLossArgument}

The photino flux constitutes a source of luminosity loss for the star. This energy depletion must be lower than our precision measurements of the sun's luminosity. The best constraints come from helioseismological and neutrino emission observations. Following \cite{krief2016solar}, we can impose a bound on the photino luminosity,

\begin{equation}
    L_\Lambda \leq 0.003  \ L_\odot,
\end{equation}
which implies the following bound on the strength of the fermionic background,
\begin{equation}
\label{FermionicBound}
    \boxed{|\psi|^2 \leq 6.33 \times 10^{-34} \ eV.}
\end{equation}
The order of magnitude found above for the fermion condensate present in the LSV background supermultiplet is compatible with the scale of the CFJ as reported in the work of Ref. \cite{kostelecky2011data}.
\section{Concluding Comments}
\label{Conclusions}

In the approximation we have adopted to derive the mixing kinetic matrix in \eqref{KineticMixingMatrix}, we have considered, for simplicity, a particular situation where the following bilinear contributions to the dispersion relations of the photino are taken to be vanishing , namely, $\Bar{\Psi}\Psi=\Bar{\Psi}\gamma_5\Psi=\Bar{\Psi}\gamma_{\mu}\gamma_5\Psi=0$. These bilinears induce new effects which can be transmitted to the photon through the mixing terms. In particular, the bilinears which contain $\gamma_5$-matrices are parity-violating and could generate birefringent effects on the photon dispersion relation. This effect shall  be exploited in a forthcoming paper. 

Finally, we would like to emphasize how two different paths beyond the SM, i.e. the LSV and the SUSY, provides a new phenomenology in Physics. It is well-know that Supersymmetry provides many good candidates to dark matter, i.e. superpartners with no electric charge, and a diversity of different bounds on the so-called Minimal Supersymmetric Standard Model (MSSM) are investigated \cite{heinemeyer2022dark}. Adding LSV effects, in the approach we adopt \cite{Alvaro2004remarks}, opens up a new portal to the dark sector through the mixing term induced by the fermionic background. In a work in progress, we shall also present new phenomenological paths in a non-Abelian CFJ-type scenario \cite{AspectsofGaugeGauginoMixing}. As already mentioned, the gauge-bosons and the gauginos acquire (different) masses due to the presence of the LSV backgrounds that break SUSY. Since supersymmetric non-Abelian gauge theories allow for the coupling of gauginos to gauge fields, there is the possibility that the gauge-boson decay into two gauginos takes place at the tree-level. We prospect to inspect the physical consequences of these features in a near future.

\bibliographystyle{plain}
\bibliography{referencias}  %%% Uncomment this line and comment out the ``thebibliography'' section below to use the external .bib file (using bibtex) .





% \end{thebibliography}

\end{multicols}
\end{document}